\documentclass[article,aps]{revtex4}

\usepackage{graphicx}
\usepackage{bm}
\usepackage{color}

\newcommand{\be}{\begin{equation}}
\newcommand{\ee}{\end{equation}}
\newcommand{\bea}{\vspace{0.25cm}\begin{eqnarray}}
\newcommand{\eea}{\end{eqnarray}}

\def\PLA{{Phys. Lett.}  A }
\def\PRL{{Phys. Rev. Lett.} }
\def\PRA{{Phys. Rev.} A }

\begin{document}

\title{ Limit quantum efficiency for violation of
Clauser-Horne Inequality for qutrits}

\author{ M. Genovese  }

\email{genovese@ien.it}

 \affiliation{  Istituto Elettrotecnico
Nazionale Galileo Ferraris,  Strada delle Cacce 91, 10135 Torino,
Italy }

\begin{abstract}

In this paper we  present the results of numerical calculations
about the minimal value  of detection efficiency for violating the
Clauser - Horne inequality  for qutrits.  Our results show how the
use of non-maximally entangled states largely improves this limit
respect to maximally entangled ones. A stronger resistance to
noise is also found.

\end{abstract}


 \maketitle

\section{Introduction}

Quantum Mechanics presents various specific properties that
strongly differentiate it from Classical Mechanics. In the last
years these characteristic properties have been used to develop a
new research field called Quantum Information \cite{NC}, devoted
to study codification, elaboration and transmission of information
by means of quantum states with promising technological
applications. One of the more relevant properties for these
realizations is entanglement, namely the existence of quantum
correlations that cannot be reproduced by any realistic
(classical) theory based on local observables \cite{Bell}. Besides
the relevance for Quantum Information, these correlations have
allowed various experimental tests of Standard quantum Mechanics
against Local Realistic Theories \cite{Aul,MG}. In particular
various experiments have tested Bell inequalities
\cite{pdc,nos,pdc2} giving strong indications against local
realism, even if no conclusive experiment was possible due to low
detection efficiency (on the other hand locality loophole has been
closed \cite{pdc2}).

In the last years it is emerged that the use of higher dimension
Hilbert spaces ($d>2$), qudits, instead of the traditional $d=2$
ones, qubits, can be of interest in various applications to
Foundations of Quantum Mechanics and Quantum Information.

For example, it has been shown that quantum communication based on
qudits presents a higher security than the one with traditional
qubit schemes \cite{crypqun}.

Furthermore, it has also been shown that a larger violation of
Bell-like inequalities is expected for qudits
\cite{Massar1,Massar2,Massar3,Kas,Kas2,Kas3,Gisen}.

These two results have then be related in Ref. \cite{Acin}.

More in details, concerning  Bell inequalities, various studies
were addressed to understand the limit quantum efficiency for a
loophole-free test of local realism (LR) and the resistance to
noise.

For example, in Ref. \cite{Massar1} Bell inequalities with
enhanced resistance to detector inefficiency were investigated.
This is of particular interest since the loophole due to low
detection efficiency $\eta$ of the detection apparatuses is the
last unsolved problem for  a conclusive test of local realism
\cite{Aul,MG} \footnote{Incidentally, it must be noticed that a
recent experiment \cite{Win} based on the use of Be ions has
reached very high detection efficiencies (around 98 \%), largely sufficient for closing detection loophole,
but in this case not only space
like separation required for closing locality loophole was not satisfied, but the
two subsystems (the two ions) were even not really separated
during the measurement. Therefore,
this experiment cannot be considered a real
implementation of a loophole free test of Bell
inequalities, even if it represents a relevant progress in this
sense (see \cite{prep} for a general review on these problems). }.
The result was that the limit for the smallest detection
efficiency $\eta^*$ necessary for a loophole free test of LR,
decreases for $d>2$ maximally entangled states of a $1-2 \%$
respect to the value $\eta^* = 82.84 \%$ for $d=2$ maximally
entangled states with 2x2 number of settings of the detection
apparatuses. In Ref. \cite{Massar2} it was then shown that for a
specific hidden variable model differences between Quantum
Mechanics (QM) and Local Realistic Theories (LRT) are observable
up to $\eta^* > { M_A + M_B -2 \over M_A M_B -1}$, where $M_A$ and
$M_B$ are the number of measurements available to the two
experimenters (usually dubbed Alice and Bob) sharing two
subsystems of a general entangled state. Finally, an asymptotic
result for large $d$ was obtained in Ref. \cite{Massar3}.

On the other hand, the resistance to noise of  some specific Bell
inequalities tested by using maximally entangled states generated
by multiport beam splitters  was investigated in Ref.
\cite{Kas,Kas2}, showing how it increases with $d$. In Ref.
\cite{Kas2} it was also shown how, for maximally entangled states,
the limit detection efficiency decreases from $0.8285$ for $d=2$
up to $0.8080$ for $d=16$ (being $0.8209$ for qutrits, $d=3$).  In
Ref. \cite{Kas3} a specific Clauser-Horne like inequality was
proposed and investigated for the previous maximally entangled
system (inequality that includes also the ones presented in \cite{Col}).

Similar results concerning the resistance to noise of LR tests
performed with qudits were obtained in Ref. \cite{Gisen} as well.

Finally,  first experimental test of Bell Inequalities with
qutrits have been recently realized  \cite{Vaz,bou,Gisen2}.

Considering the large interest both in the fields of Quantum
Information \cite{NC} and of Foundations of Quantum Mechanics
\cite{Aul} for a characterization of violations of Local Realism
in $d>2$ Hilbert spaces, in this paper we numerically analyze the
detection efficiency limit for non-maximally entangled states for
the Clauser-Horne inequality proposed in Ref. \cite{Kas3}. The
interest for studying non-maximally entangled states is a
consequence of the fact that for qubits it was shown \cite{eb}
that a smaller detection efficiency, $66.7 \%$, is needed for
closing the detection loophole than the one required for maximally
entangled states, $82.8 \%$ (for an experimental test of $d=2$
Clauser-Horne inequality with non-maximally entangled states see
Ref. \cite{nos}).

 In order to give a first hint to the effect of using non-maximally entangled states,
 we will consider two examples:
one is given by entangled systems built by multiport
beam-splitters (interferometers), a second is given by an
entangled state generated by using  biphoton states \cite{moscow}.
Both these qutrits states have already been realized with photons
\cite{moscow2,Gisen2} (for the first also entanglement has been
obtained \cite{Gisen2}) and, therefore, our results, showing a
relevant improvement by using non-maximally entangled states, are
of a large relevance for future experimental tests of LRT against
QM and applications to quantum information.

\section{Clauser-Horne inequality for qutrits}

The Clauser-Horne inequality, valid for LRT, introduced in Ref.
\cite{Kas3} is

\bea CH= P^{11}(2,1) + P^{12}(2,1) - P^{21}(2,1)  + P^{22}(2,1) +
\cr P^{11}(1,2) + P^{12}(1,2) - P^{21}(1,2)  + P^{22}(1,2) +\cr
P^{11}(2,2) + P^{12}(1,1) - P^{21}(2,2)  + P^{22}(2,1) - \cr
\left( P^1_1(1) + P^1_1(2)+P^2_2(1)+P^2_2(2) \right) <= 0 \,\,
\label{CH}
 \eea

where $P^{ij}(k,l)$ denotes the joint probability for Alice
measuring in the basis $i$ and Bob in the basis $j$ ($i,j = 1,2$)
a thricotomic observable to obtain the result $k$ and $l$
respectively ($k,l=1,2,3$), whilst $P^i_n(k)$ denotes the single
measurement probability of Alice ($n=1$) or Bob ($n=2$) to obtain
the result $k$ when the basis $i$ is used \footnote{It is
interesting to notice that this inequality even if valid for
qutrits does not involve the third outcome of the measurement
\cite{Kas3}.}.

While $CH <= 0$ for every LRT, in Ref. \cite{Kas3} was shown that
this inequality is violated (with a maximal violation $CH =
0.29098$) by a qutrit maximally entangled state.

This state is obtained by applying a tritter (unbiased 6-port
beamsplitter) plus three phase shifts, which altogether are
described by the unitary transformation
 \be U_{kl} = {1 \over \sqrt{3} }
exp (i 2 \pi/3 (k-1) (l-1)) \cdot exp ( i \phi _l ) \label{U} ,
\ee

 to the state

\be \Psi = {1 \over \sqrt{3}} (| 1 \rangle_A | 1 \rangle _B + | 2
\rangle _A | 2 \rangle _B + | 3 \rangle _A | 3 \rangle _B   )
\label{psim} \ee

where the states $| i \rangle_{A,B}$ describe the i-th basis state
of the qutrit, respectively sent to Alice and Bob.

The measurement is described by the projection in one of these
states $| i \rangle$. Thus the quantum probabilities are:

\bea P^{ij}(k,l) = |\langle k_A | \langle l_B| U_A(\vec{\phi_i})
U_B(\vec{\theta_j}) | \Psi \rangle |^2, \\
P^i_n(k) = | \langle k_n | U_n (\vec{\phi_i}) | \Psi \rangle |^2
 \eea

 The result of a violation of the Clauser-Horne inequality, $CH = 0.29098$, corresponds to a limit for the
minimal detection efficiency of $\eta^* = 0.8209$ in perfect
agreement with Ref. \cite{Kas2}.

In the following we will reconsider this violation for
non-maximally entangled states with the purpose of verifying if
also in this case, as in the qubits one \cite{eb}, a substantial
improvement on the value of the minimal detection efficiency
appears.

\section{Numerical results with non-maximally entangled states}

Let us begin by reconsidering the violation of inequality \ref{CH}
in the case of a non-maximally entangled state obtained by the
transformation \ref{U} applied to the state (a,b are chosen to be
real):

\be \Psi = {1 \over \sqrt{1 + a^2 + b^2 }} \left(| 1 \rangle _A |
1 \rangle _B + a | 2 \rangle _A | 2 \rangle _B + b | 3 \rangle _A
| 3 \rangle _B  \right ) \,\,\,\, \label{psinm} \ee

This state could be easily obtained with a simple (unbalanced
tritter) modification of the experimental scheme of Ref.
\cite{Gisen2}.

By using analytic expressions for coincidence probabilities
deriving by \ref{psinm} we perform a numerical maximization (by
using the software Mathematica) on the phases $\vec{\phi}$ and
$\vec{\theta}$ and on the parameters $a,b$.

Our numerical simulation
 shows that the limit for the detection
efficiency $\eta^* = 0.8209$ for maximally entangled states, can
be lowered to $\eta^* = 0.8139$ for non-maximal entanglement. This
result is not particularly large, even if it is analogous to the
ones obtained increasing the dimension $d$ of the Hilbert space
and/or the number of measurements reported in Ref.
\cite{Kas2,Massar1}.

In order to give an idea of the region where the inequality
\ref{CH} is violated,  in Fig.1 we plot the value of the quantity
$CH$, Eq. \ref{CH}, around one maximum in function of the two
parameters $a,b$  for a detection efficiency $\eta = 0.85$.

\begin{figure}
   \begin{center}
   \begin{tabular}{c}
   \includegraphics[height=7cm]{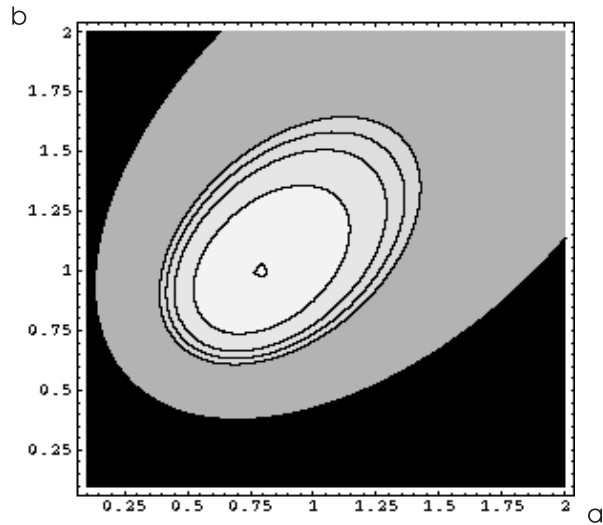}
   \end{tabular}
   \end{center}
   \caption[example]
   { \label{fig:1}

Contour plot of the quantity $CH$ (see Eq. \ref{CH}), calculated
with the qutrit state generated by the tritter scheme, in the
plane a (abscissa) - b (ordinate) in a region around a maximum for
a detection efficiency $\eta = 0.85$. Contour lines are at
-0.1,0,0.007,0.015,0.03,0.05.}

   \end{figure}

   \begin{figure}
   \begin{center}
   \begin{tabular}{c}
   \includegraphics[height=7cm]{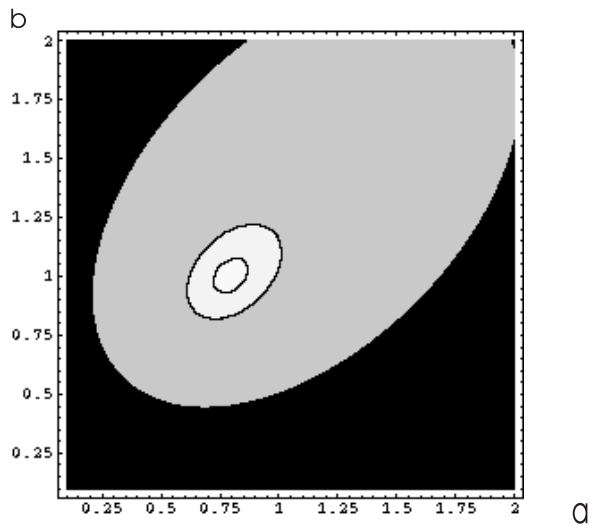}
   \end{tabular}
   \end{center}
   \caption[example]
   { \label{fig:2}
Contour plot of the quantity $CH$ (see Eq. \ref{CH}), calculated
with the qutrit state generated by the tritter scheme, in the
plane  $a-b$ in a region around a maximum for a detection
efficiency $\eta = 0.82$. Contour lines are at -0.1,0,0.007.}
   \end{figure}

In Fig.2 the same contour plot is shown for the case $\eta =
0.82$. It is evident that the point $a=1,b=1$ (maximal
entanglement), which is eccentric from the maximum but still in a
positive region for $\eta = 0.85$, becomes just outside the
positive region for $\eta = 0.82$.

The relatively small numerical relevance of using non-maximal
entangled states can be due to the specific choice of the
entangled state. In particular, in this case the single detector
probabilities $P^i_n(k)$ are independent on the parameters $a$ and
$b$ and equal to $1/3$. Thus the last line of inequality \ref{CH}
is constant, giving a relatively large negative contribution
$-4/3$ (we will see later than in other cases the contribution of
this part can be reduced). The use of other states can change this
situation and eventually improve the result about the limit
detection efficiency.

Let us analyze more in details this point, beginning with the
qubit case.
 For the $d=2$ non-maximally entangled state:

\be \vert \Psi \rangle ={\vert 0 \rangle \vert 0 \rangle + a \vert
1 \rangle \vert 1 \rangle \over \sqrt{1 + \vert a \vert ^2}}
\label{PsiH} \ee

the original Clauser-Horne sum is

\bea CH(d=2)=p(\theta _{1},\theta _{2})-p(\theta _{1},\theta
_{2}^{\prime })+ \cr
p(\theta _{1}^{\prime },\theta _{2})+
 p(\theta _{1}^{\prime },\theta _{2}^{\prime })-  p(\theta
_{1}^{\prime })-p(\theta _{2}) \,\, \label{eq:CH} \eea

 which is strictly negative
for every local realistic theory. In (\ref{eq:CH}), $p(\theta
_{1},\theta _{2})$ is the probability of coincidences between
channels 1 and 2 when $\theta _{1}$ and $\theta _{2}$ selections
are performed, $p(\theta _{i})$ are single detector count
probabilities corresponding to the selection $\theta_i$. In Fig. 3
we report the contour plot for $CH$ in the qubit case in the plane
$\eta - a$

If one considers a maximally entangled state ($a=1$) for $\eta=1$
the inequality is violated by a quantity $CH= 0.2071$. The
(negative) contribution from the single particle probabilities,
$CH_{single}$ is $82.85 \%$ of the joint probabilities one,
$CH_{joint}$.

When $\eta$ is decreased, for example to $0.85$, for the maximally
entangled state the violation is reduced to $CH=0.0221$ and now
the contribution $CH_{single}$ is the $97.4 \%$ of $CH_{joint}$.
If the entanglement parameter $a$ is varied as well the maximal
violation is $CH (\eta = 0.85) = 0.0496$ for $a=0.608$. In this
case the contribution of $CH_{single}$ respect to $CH_{joint}$ is
reduced to $90.76 \%$.

\begin{figure}
   \begin{center}
   \begin{tabular}{c}
   \includegraphics[height=7cm]{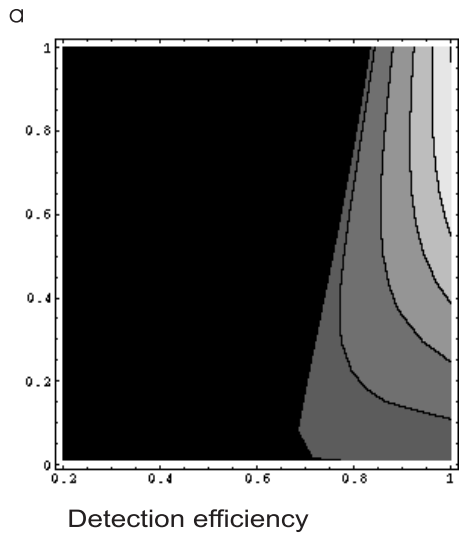}
   \end{tabular}
   \end{center}
   \caption[example]
   { \label{fig:f} Contour plot of the quantity $CH(d=2)$ (see Eq. \ref{eq:CH})
   in the plane with $a$ (non maximally entanglement parameter, see
the text for the definition) as ordinate and $\eta$ (total
detection efficiency) as abscissa. The leftmost region (in black)
corresponds to the region where no detection loophole free test of
Bell inequalities can be performed.
 The contour lines are at 0,0.01,0.05,0.1,0.15,0.2.
One can observe how the lowest quantum efficiency for having
$CH>0$ is 0.667 for $a \approx  0.15$. }
  \end{figure}

On the other hand for the qutrit case just considered, the ratio
$CH_{single} / CH_{joint}$ is  $82.1 \%$  for a maximally
entangled state when $\eta =1$ and becomes $CH_{single} /
CH_{joint}= 0.9657 $ for $\eta = 0.85$ ($CH = 0.0402$). In this
second case, for a non-maximally entangled state it only reduces
to $CH_{single} / CH_{joint}= 0.9575$ ($CH=0.05033$).

In order to better investigate this point, we have therefore
considered as a second example the case where qutrit basis is
realized by using degenerate biphotons \cite{moscow,moscow2}

\bea | 1 \rangle = | HH \rangle \cr | 2 \rangle = | HV \rangle \cr
| 3 \rangle = | VV \rangle \eea

where $H$ and $V$ denote horizontal and vertical polarization of
photons respectively. In this specific case the two observable
pertains to two independent photons and therefore the detection
efficiency for the single state, here a biphoton, corresponds to
the product of the quantum efficiencies of two photon-detectors.
Nevertheless, since we are interested in a general theoretical
discussion about how large can be the effect of using
non-maximally entangled states for qutrits case (in a general
abstract Hilbert space), in the following we discuss the results
in term of the detection efficiency for the state (of course, the
square root of it immediately gives the single photon-detector
one).

The measurement, performed on the non-maximally entangled state
\ref{psinm}, is both for Alice and Bob, represented by a
polarization selection on a chosen basis at a certain angle
$\theta$ from the horizontal direction. The results of this
measurement will be:

i) both the photons passing the selection (e.g. both up if  a
polarizing beam splitter is used),

ii)  both not passing it (e.g. both down)

iii) one passing and the other not the selection (e.g. one up and
one down)

These cases are respectively represented by the projectors:

\bea P_1 =  | 2_{\theta} \rangle \langle 2_{\theta} | \cr P_2 =  |
2_{\theta + \pi/2} \rangle \langle 2_{\theta + \pi/2} | \cr P_3= |
1_{\theta} 1_{\theta + \pi/2} \rangle \langle 1_{\theta} 1_{\theta
+ \pi/2} | \cr\eea

where $| n_{\theta } \rangle$ denotes the state of $n$ photons
with polarization along an axis forming an angle $\theta$ with the
horizontal direction.

Experimentally this state could be generated with the techniques
presented in Ref. \cite{multi} for entangling in polarization four
(or more) photons states.

 In our numerical simulation we have initially chosen as the measurement results
1,2 in the inequality \ref{CH} the first two former cases
($P_1,P_2$). We have then maximized the quantity $CH$ on the Alice
and Bob bases ($\theta_A^1,\theta_A^2,\theta_B^1,\theta_B^2$) and
on the entanglement parameters $a,b$.

Our numerical results show that for a maximally entangled state
($a=b=1$) the maximal violation is $CH=0.1765$ corresponding to a
limit of the detection efficiency  $\eta^* = 0.8835$. On the other
hand, when a generic non maximal entangled state is chosen, the
result strongly improves reaching as limit for the detection
efficiency  $\eta^* = 0.76$.

For the sake of exemplification, in Fig.4 we plot the value of the
quantity $CH$, Eq. \ref{CH}, in function of the two parameters
$a,b$  for a detection efficiency $\eta = 0.85$.

\begin{figure}
   \begin{center}
   \begin{tabular}{c}
   \includegraphics[height=7cm]{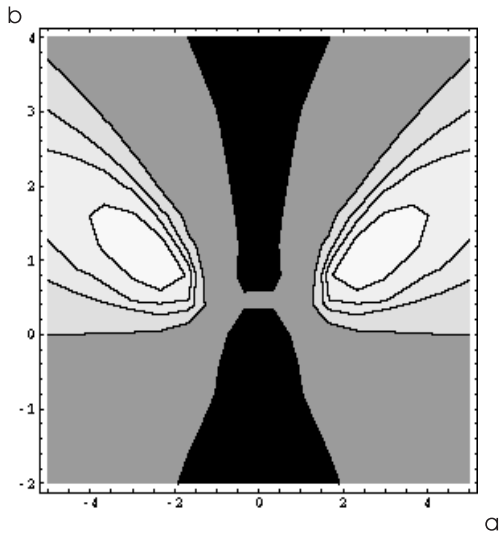}
   \end{tabular}
   \end{center}
   \caption[example]
   { \label{fig:bpq}
Contour plot of the quantity $CH$ (see Eq. \ref{CH}), calculated
for biphoton qutrits, in the plane a (abscissa) - b (ordinate)
for a detection efficiency $\eta = 0.85$. Contour lines are at
CH=-0.1,0,0.01,0.015,0.02.}
   \end{figure}

The effect of using non-maximally entangled states is therefore
rather large, even if still smaller than the effect for qubits
(where it is lowered \cite{eb} from $\eta^* = 0.828$ to $\eta^*
=0.667$). In particular, the effect is larger here than for
qutrits generated by using tritters.

This last result is related to the fact that in this second case
the contribution from the single probabilities decreases
relevantly for a non-maximally entangled state. Our results are
that for a maximally entangled state this contribution increases
from $88.31 \%$ when $\eta = 1$ to $98.12 \%$ when $\eta = 0.9$
($CH = 0.02297$). On the other hand, leaving $a,b$ free to vary we
obtain at $\eta =1$  a maximal violation $CH = 0.1851$ with
$CH_{single} / CH_{joint} = 0.867$. At $\eta = 0.9$ single
probabilities contribution becomes $91.88 \%$ of the joint
probabilities one (with $CH= 0.06166$ when $a=1.820,b=1.002$
\footnote{Other minima correspond to other parameters values.} ).

Even a larger effect is obtained when the case $P_1,P_3$ is
chosen. For the maximally entangled state $a=b=-1$ we find
$CH=0.21007$ for $\eta = 1$, with a ratio $CH_{single} /
CH_{joint} = 0.8507$. On the other hand when the parameters $a,b$
are left free to vary, we obtain a much larger violation, $CH=
0.31607$, corresponding to a reduction of the ratio $CH_{single} /
CH_{joint} $ to $0.7610$ ($a= -1.37,b=-0.607$). For $\eta = 0.9$
we have $CH= 0.06256$ with $CH_{single} / CH_{joint} = 0.945$ for
the maximally entangled state, whilst $CH = 0.1686$, $CH_{single}
/ CH_{joint} = 0.8334$ for a non-maximally entangled one. Finally,
the limit for violating the Clauser-Horne inequality is  $\eta^* =
0.8505$ for maximal entanglement, reduced to $\eta^* = 0.7413$ for
$a= -1.755,b=-0.572$. For the sake of comparison, the best results
for qubits, tritter and biphoton qutrits are shown in table 1.

In the last case  also the resistance to noise is
increased respect to the result of Ref. \cite{Kas3} for maximally
entangled states, since the value of $CH$ is increased.
If, in analogy to Ref. \cite{Kas3}, one considers
the mixed state ($0 \le F \le 1$)
 \be \rho = (1-F) | \Psi \rangle \langle \Psi | + F
\rho_{noise} \ee where $\rho_{noise} $ is a diagonal matrix with
entries equal to $1/9$ the threshold value of $F$ for violating
the Clauser-Horne inequality  is $F_{th} = 0.3216$ to be compared
with $F_{th} = 0.30385$ obtained in Ref. \cite{Kas3,Col,Kasq}. Incidentally,
the result that noise threshold is higher for non-maximally entangled
states follows also from the results of Ref. \cite{ac}, from which
one obtains $F_{th} = 0.31471$.

 These results further show the large effect on the violation
of Clauser - Horne inequality obtained by using non-maximally
entangled states. Since the maximum has been evaluated by a
numerical maximization even larger effects cannot be completely
excluded.

\vskip 0.2cm
\begin{tabular}{|c|c|c|}
  \hline
  & Maximally entangled states & Non-maximally entangled states \\
  \hline
  qubits & 0.828 & 0.667 \\
  tritter qutrits & 0.8209 & 0.8139 \\
  biphoton qutrits & 0.8505 & 0.7413 \\
  \hline
\end{tabular}

Table 1: limit on the detection efficiency for a certain state
necessary for obtaining a detection loophole free experiment.

\section{Conclusions}

In conclusion in this paper we have presented the results of
numerical simulations about the minimal value  of detection
efficiency for violating the Clauser - Horne inequality  for
qutrits proposed in Ref. \cite{Kas3}. In particular we have
considered the case of entangled qutrits generated by means of a
tritter scheme (as in Ref. \cite{Kas3}) and of qutrits based on
biphotons (as suggested in \cite{moscow,moscow2}).

Our results show that, in analogy to the case of qubits \cite{eb},
the use of non-maximally entangled states may largely improve this
limit. Nevertheless, in these specific numerical examples, the
effect remains smaller than in qubit case. The difference among
the cases considered here show how the specific numerical results
largely depend on the choice of states and measurements.
Altogether, our results give a first indication of the interest of
using non-maximally entangled states also for $d>2$ Hilbert
spaces, pointing out the relevance of further theoretical and
experimental studies in this sense.

\vskip 0.5cm

{\bf Acknowledgments}

I acknowledge the support of MIUR (FIRB RBAU01L5AZ-002) and
Regione Piemonte.

I thank anonymous referee for having pointed out to my attention
that from Ref. \cite{ac} one obtains for non-maximally entangled
states the threshold value $F_{th} = 0.31471$, now reported in the
text.

 \vskip 3cm


\end{document}